\documentstyle[emulateapj]{article}
\begin{document}
\lefthead{IRWIN ET AL.}
\righthead{NGC~4697}

\slugcomment{Astrophysical Journal, accepted}

\title{The X-ray Faint Early-Type Galaxy NGC~4697}

\author{Jimmy A. Irwin\altaffilmark{1,3}, Craig L. Sarazin\altaffilmark{2},
and Joel N. Bregman\altaffilmark{1}}

\altaffiltext{1}{Department of Astronomy, University of Michigan, \\
Ann Arbor, MI 48109-1090 \\
E-mail: jirwin@astro.lsa.umich.edu, jbregman@umich.edu}

\altaffiltext{2}{Department of Astronomy, University of Virginia,
P.O. Box 3818, Charlottesville, VA 22903-0818;
cls7i@virginia.edu}

\altaffiltext{3}{Chandra Fellow.}

\begin{abstract}
We analyze archival {\it ROSAT} HRI, {\it ROSAT} PSPC, and {\it ASCA} data
of the X-ray faint early-type galaxy NGC~4697. The joint {\it ROSAT} PSPC +
{\it ASCA} spectrum is fit by a two-component thermal model, a MEKAL model
with $kT_{MEKAL}=0.26^{+0.04}_{-0.03}$ keV with low metallicity and a
bremsstrahlung model with $kT_{BREM}=5.2^{+3.0}_{-1.6}$ keV. A similar model
was found to fit the spectra of another faint early-type galaxy (NGC~4382) and
the bulge of M31. We interpret this soft emission as a combination of
emission from a soft component of low mass X-ray binaries (LMXBs)
and from a low temperature interstellar medium,
although the relative contributions of the two components could not be
determined. Twelve point sources were identified within $4^{\prime}$ of
NGC~4697, of which 11 are most likely LMXBs associated with the
galaxy. The soft X-ray colors of four of the LMXBs in NGC~4697 support the
claim that LMXBs possess a soft spectral component. Finally, we present a
simulation of what we believe the {\it Chandra} data of NGC~4697 will look
like.

\end{abstract}


\section{Introduction} \label{sec:intro}

It is well-established that early-type galaxies exhibit a large range
of X-ray--to--optical luminosity ratios, $L_X/L_B$.
Although there is a strong correlation between the X-ray and optical
luminosities of elliptical and S0 galaxies ($L_X \propto L_B^{1.7-3.0}$;
Canizares, Fabbiano, \& Trinchieri 1987; White \& Davis 1997;
Brown \& Bregman 1998), a large dispersion exists in this relation.
Two galaxies with similar blue luminosities might have X-ray luminosities
that differ by as much as a factor of 100
(Canizares et al.\ 1987; Fabbiano, Kim, \& Trinchieri 1992;
Brown \& Bregman 1998).

Hot ($\sim10^7$ K) gas is believed to be responsible for the bulk of the
X-ray emission in high $L_X/L_B$ galaxies
(e.g., Forman, Jones \& Tucker 1985). Another possible source of X-ray emission
in early-type galaxies is from discrete stellar sources, primarily low-mass
X-ray binaries (LMXBs). Estimates for the stellar $L_X/L_B$ value are
uncertain and vary by a factor of ten among various studies
(Forman et al.\ 1985; Canizares et al.\ 1987). Although stellar X-ray sources
are not expected to contribute significantly to the 0.1--2.0 keV X-ray
emission in gas-rich, X-ray bright galaxies, this claim cannot be made with
any certainty for X-ray faint (low $L_X/L_B$) galaxies. In these
galaxies, it is possible that stellar X-ray sources are the dominant X-ray
emission mechanism. For this to occur, the gas lost from stellar mass loss
must be removed from the galaxy by galactic winds, by ram pressure stripping
from ambient intracluster or intragroup gas, or possibly both (see, e.g.,
Ciotti et al.\ 1991; Mathews \& Brighenti 1998).

Evidence is mounting that LMXB emission is detectable in nearly all early-type
galaxies. Matsumoto et al.\ (1997) found a hard ($\sim$5--10 keV) spectral
component in 11 of 12 early-type galaxies observed with {\it ASCA}, which
scaled roughly with the optical luminosity of the galaxy (note, however, that
Buote \& Fabian (1998) found no formal need for a hard component in the
X-ray brightest galaxies of their {\it ASCA} sample). The luminosity
of this hard component was small compared to the soft, gaseous component
in the X-ray brightest galaxies, but increased in importance as
$L_X/L_B$ decreased. However, the {\it ASCA} observations did not
resolve this component into discrete sources because of the insufficient
spatial resolution of the instrument.

Previous studies have found that the spectra of the X-ray faintest
early-type galaxies are significantly different than the spectra of
gas-rich X-ray bright early-type galaxies. {\it Einstein} data revealed
that the X-ray faintest galaxies exhibited a strong very soft X-ray excess
(Kim et al.\ 1992). Subsequent {\it ROSAT} PSPC and {\it ASCA} studies found
that the spectra of these galaxies were best fit with a two component model
consisting of the previously mentioned hard $\sim$5--10 keV component
(generally attributed to the integrated LMXB emission), and a very
soft $\sim$0.3 keV component, whose origin is uncertain
(Fabbiano, Kim, \& Trinchieri 1994; Pellegrini 1994; Kim et al.\ 1996).

One recently proposed solution to the very soft X-ray excess problem in X-ray
faint galaxies is that it might result from the very same collection
of LMXBs responsible for the hard emission (Irwin \& Sarazin 1998a,b).
Little is known about the very soft X-ray properties of LMXBs since
nearly all Galactic examples lie in directions of high Galactic hydrogen
column densities, so their soft X-ray emission is heavily absorbed.
However, two nearby Galactic LMXBs that lie in directions of low Galactic
hydrogen column density (Her X-1 and MS1603+2600) exhibit significant
very soft X-ray emission (Vrtilek et al.\ 1994; Hakala et al.\ 1998).
But perhaps the strongest evidence comes from the bulge of M31. M31 is
close enough that most ($\ga$75\%) of its bulge X-ray emission was
resolved into point sources with the {\it ROSAT} PSPC and HRI
(Supper et al.\ 1997; Primini, Forman, \& Jones 1993), a majority of
which are probably LMXBs. Both individually and cumulatively, these
point sources in the bulge of M31 have X-ray spectral properties very similar
to the integrated emission from X-ray faint early-type galaxies in the
{\it ROSAT} band (Irwin \& Sarazin 1998a,b). A joint {\it ROSAT} PSPC +
{\it ASCA} study found the X-ray spectrum of the bulge of M31 to be almost
identical to that of the X-ray faint early-type galaxy NGC~4382 over the
0.2--10 keV band (Irwin \& Bregman 1999a; Kim et al.\ 1996).
The bulge of the Sa galaxy NGC~1291
also exhibits a very soft component and has X-ray colors identical to
the bulge of M31, although it is unresolved with
{\it ROSAT} because of its distance. In both bulges, the measured
$L_X/L_B$ value is comparable to those of the X-ray faintest early-type
galaxies. This suggests that in addition to providing the required spectral
X-ray characteristics, LMXBs are luminous and/or numerous enough
to produce the required amount of X-ray emission in the X-ray faintest
early-type galaxies, as well as Sa spiral bulges. No additional very soft
X-ray emission source seems to be required.

In this {\it Paper}, we analyze long {\it ROSAT} PSPC, {\it ROSAT} HRI, and
{\it ASCA} GIS and SIS observations of the X-ray faint elliptical galaxy
NGC~4697.
At a distance of 15.9 Mpc (Faber et. al.\ 1989; assuming a Hubble constant
of 50 km s$^{-1}$ Mpc$^{-1}$), NGC~4697 is one of the closest
normal early-type galaxies. Previous work has indicated that this galaxy
has a below average but not extremely low $L_X/L_B$ value, and {\it ROSAT}
band colors typical
of those of other X-ray faint early-type galaxies (Irwin \& Sarazin 1998b).
Jointly fitting the {\it ROSAT} PSPC and {\it ASCA} spectra allows us to place
useful constraints on both the hard and soft emission, something that is not
possible using each instrument separately. In addition, NGC~4697 was one
of the very few X-ray faint early-type galaxies observed at length with the
{\it ROSAT} HRI. We analyze this HRI image of NGC~4697 to resolve as much of
the X-ray emission as possible into discrete sources. We also present a
{\it Chandra} simulation of what the X-ray emission of NGC~4697 might look
like.

\begin{figure*}[htb]
\vskip4.4truein
\includegraphics{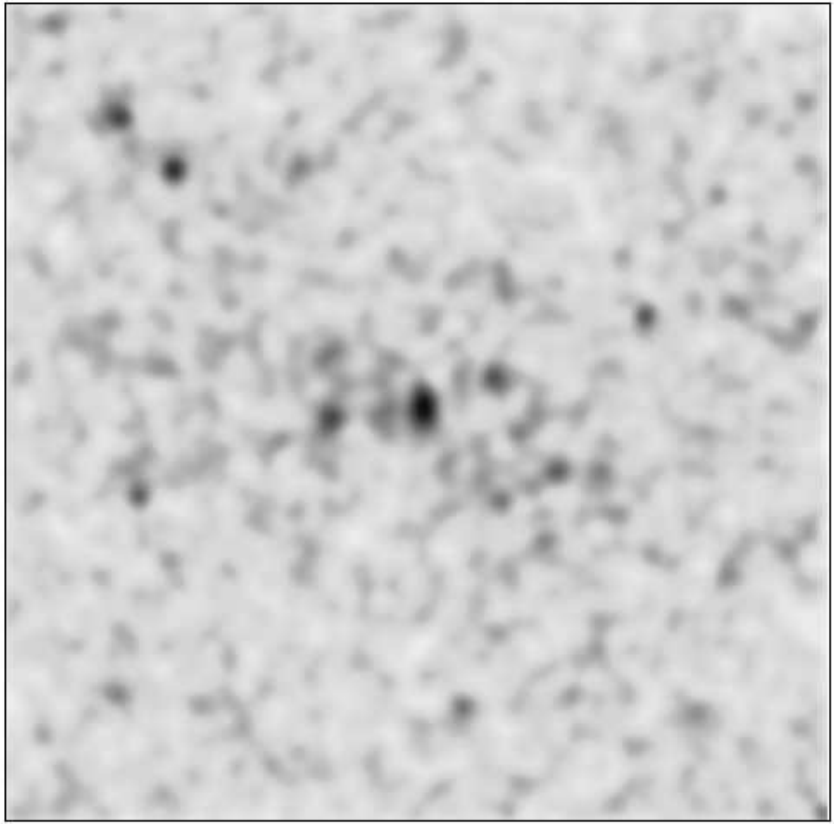}
\caption[]{
The inner $8^{\prime} \times 8^{\prime}$ of the {\it ROSAT} HRI
image of NGC~4697. The HRI map has been smoothed with a Gaussian of
$5^{\prime\prime}$.
\label{fig:hri}}
\end{figure*}

\section{{\it ROSAT} PSPC and {\it ASCA} Data Reduction} \label{sec:data}

 From the HEASARC archive we have extracted long {\it ROSAT} PSPC (RP600262A02)
and {\it ASCA} GIS and SIS (62014000) observations of NGC~4697. The PSPC data
were filtered such that all data with a Master Veto Rate below
170 counts s$^{-1}$ were excluded from the data, yielding an observation time
of 36,856 seconds. The {\it ASCA} data were screened using the standard
screening criteria applied to all the archival data (Revision 2 processing).
See Ohashi et al.\ (1996) and Makishima et al.\ (1996) for a description of
the {\it ASCA} instruments. The total GIS and SIS exposure times were 56,698
seconds and 42,138 seconds, respectively. The SIS observation was taken in
2-CCD mode.

Spectra from a 4$^{\prime}$ radius were extracted from the PSPC, GIS, and
SIS data. An extraction area of this size provided a good compromise
between the minimum suggested extraction region for SIS data and achieving a
reasonable signal-to-noise ratio for the PSPC data. For the PSPC we have chosen
background from a source-free annular region $30^{\prime}-40^{\prime}$ in
extent corrected for vignetting. For the {\it ASCA} data, background was
extracted from the deep blank sky data provided by the {\it ASCA} Guest
Guest Observer Facility. We used the same region filter to extract the
background as we did the data, so that both background and data were affected
by the detector response in the same manner.

For the PSPC data, only energy channels between 0.2--2.0 keV were included
in the fits, and for the {\it ASCA} data we used energy channels between
0.8--10 keV. The energy channels were regrouped so that each channel contained
at least 25 counts so that the $\chi^2$-test is a valid indicator of
goodness of fit. For all the joint fits, we linked the normalizations of the
PSPC and GIS, but let the SIS normalization vary to account for the fact that
some of the emission that fell on interchip boundaries of the SIS was lost.

\section{Spectral Modeling and Luminosities of the Global Spectrum}
\label{sec:spectral}

Using XSPEC Version 10.0, we first attempted to fit a single component thermal
model to the PSPC and {\it ASCA} data separately. When the PSPC spectrum was fit
alone with a MEKAL model with a variable absorption component, a good fit was
obtained ($\chi_{\nu}^2=1.07/58$
degrees of freedom) for $kT=0.43-0.56$ keV and $Z<0.02$ (90\% confidence level),
in good agreement with the analysis of the same data by Davis \& White (1996).
A significantly different result was found for this same model when just the
{\it ASCA} spectra were analyzed; the best-fit values were $kT=2.2-3.9$ keV
and $Z<0.20$, with $\chi_{\nu}^2=1.44/53$ degrees of freedom. Clearly,
significantly different conclusions would be drawn from the spectra if data
from only one of the satellites were analyzed.

This result illustrates the necessity of using both {\it ROSAT} PSPC and
{\it ASCA} data when analyzing the spectrum of X-ray faint early-type galaxies.
When a one-component MEKAL model with a variable absorption component was fit
to the joint PSPC+{\it ASCA} spectrum, a poor fit was obtained
($\chi_{\nu}^2=2.08/115$ degrees of freedom) with $kT=0.82$ keV and $Z<0.006$.
Following the work of Kim et al.\ (1996) and Irwin \& Bregman (1999a), we
added a bremsstrahlung model in the fit. The MEKAL+bremsstrahlung model
provided an excellent fit to the data ($\chi_{\nu}^2=0.94/113$ degrees of
freedom) with the absorption value fixed at the Galactic
line-of-sight value of $2.12 \times 10^{20}$ cm$^{-2}$ (Stark et al.\ 1992).
The MEKAL model had parameter values of $kT_{MEKAL}=0.26^{+0.04}_{-0.03}$ keV
and $Z=0.07^{+0.05}_{-0.03}$ (all errors are 90\% confidence levels for one
interesting parameter). The best-fit bremsstrahlung 
temperature was $kT_{BREM}=5.2^{+3.0}_{-1.6}$ keV. Freeing the absorption
parameter led to only a slight improvement in the fit, and the 90\%
confidence value on $N_H$ was consistent with the Galactic value.

We hesitate to attach a physical significance to this MEKAL+bremsstrahlung
model. The spectra of individual Galactic LMXBs are often fit by a
disk-blackbody+blackbody (DBB+BB) spectrum (see, e.g., Mitsuda et al.\ 1984).
The MEKAL+bremsstrahlung, however, provided a good fit to the X-ray faint
early-type galaxy NGC~4382 (Kim et al.\ 1996) and the bulge of M31
(Irwin \& Bregman 1999a; Trinchieri et al.\ 1999 showed that a
Raymond-Smith+bremsstrahlung model adequately fit the {\it BeppoSAX} LECS
spectrum of the bulge of M31), whereas the DBB+BB model did not provide a good
fit to the {\it ROSAT} PSPC + {\it ASCA} spectrum of the bulge of M31. Since we
are interested in describing the spectrum of the a collection of LMXBs in the
context of an elliptical galaxy and not of LMXBs on an individual basis, we
will continue to use a MEKAL+bremsstrahlung model as has been done in previous
studies of X-ray emission from early-type systems. We do not attempt to
justify the physical relevance of the MEKAL+bremsstrahlung model, but use it
here to quantify the strength of the soft and hard components of the X-ray
emission.

Assuming a distance of 15.9 Mpc, the 0.25--10 keV luminosities of the soft and
and hard components were $1.50 \times 10^{40}$ ergs s$^{-1}$ and
$1.83 \times 10^{40}$ ergs s$^{-1}$, respectively, in the 0.25--10 keV band.
In the {\it ROSAT}
band (0.1--2.4 keV) the luminosities were $2.36 \times 10^{40}$ ergs s$^{-1}$
and $1.04 \times 10^{40}$ ergs s$^{-1}$, respectively.

\begin{table*}[htb]
\begin{center}
\caption{\hfil HRI X-ray Sources Within $4^{\prime}$ \label{tab:sources} \hfil}
\begin{tabular}{lccccc}
\tableline
\tableline
Source&R.A.   &Decl.   &Counts&S/N&X-ray Luminosity (ergs s$^{-1}$)\\
\tableline
1 &12:48:27.13&-5:47:05.3&27.6$\pm$7.4&3.7&$7.0 \times 10^{38}$\\
2 &12:48:30.54&-5:48:34.3&17.6$\pm$6.5&2.7&$4.4 \times 10^{38}$\\
3 &12:48:32.05&-5:48:14.1&14.1$\pm$6.20&2.3&$3.6 \times 10^{38}$\\
4 &12:48:32.60&-5:48:52.3&16.2$\pm$6.4&2.5&$4.1 \times 10^{38}$\\
5 &12:48:33.03&-5:47:39.8&22.1$\pm$6.9&3.2&$5.6 \times 10^{38}$\\
6 &12:48:34.24&-5:50:51.5&17.8$\pm$6.5&2.7&$4.5 \times 10^{38}$\\
7\tablenotemark{a}&12:48:35.65&-5:48:01.3&30.5$\pm$11.3&2.7&$7.7\times10^{38}$\\
8\tablenotemark{a}&12:48:35.68&-5:47:54.3&30.0$\pm$11.3&2.7&$7.6\times10^{38}$\\
9 &12:48:39.30&-5:48:03.9&20.0$\pm$6.7&3.0&$5.0 \times 10^{38}$\\
10&12:48:45.41&-5:45:40.1&32.8$\pm$7.8&3.5&$8.3 \times 10^{38}$\\
11&12:48:46.77&-5:48:51.1&17.8$\pm$6.5&4.2&$4.5 \times 10^{38}$\\
12&12:48:47.42&-5:45:10.0&24.0$\pm$7.1&3.4&$6.0 \times 10^{38}$\\
\tableline
\end{tabular}
\end{center}
\tablenotetext{a}{Count rate and luminosity uncertain because of crowding}
\end{table*}

\section{{\it ROSAT} HRI Observation of NGC~4697} \label{sec:HRI}

\begin{figure*}[htb]
\vskip4.6truein
\includegraphics{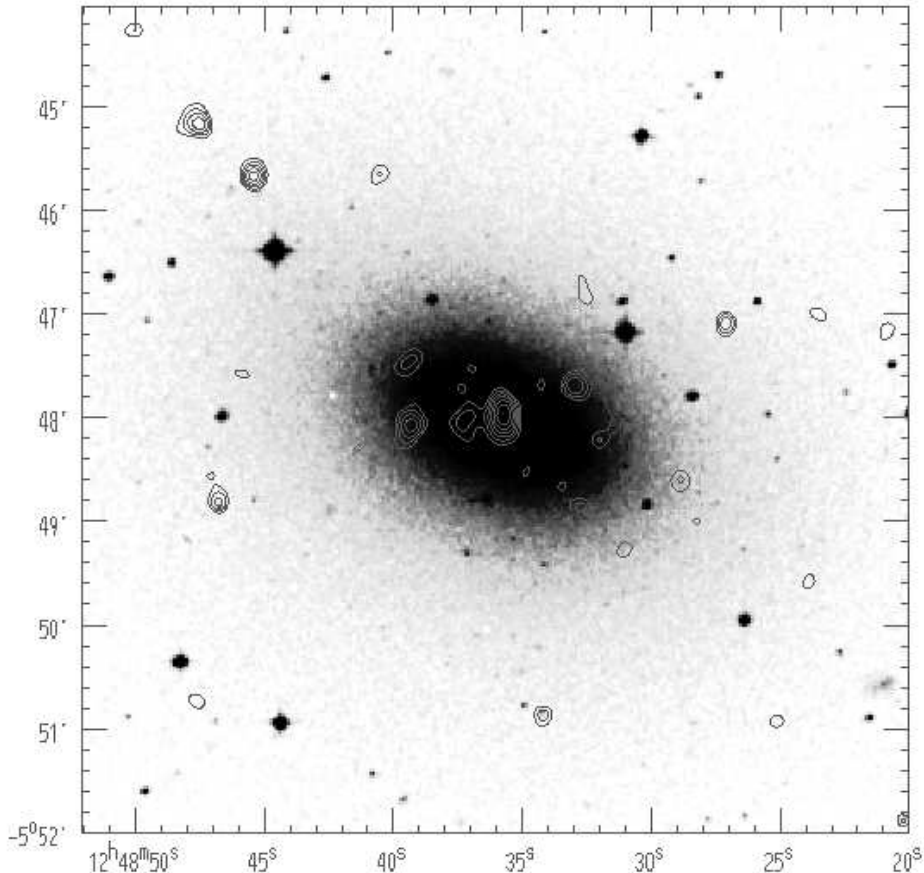}
\caption[]{
{\it ROSAT} HRI contour map overlaid the Digital Sky Survey image of NGC~4697.
The HRI map has been smoothed with a Gaussian of $5^{\prime\prime}$, and the
contours have been chosen to highlight the point sources.
\label{fig:hri_opt}}
\end{figure*}

NGC~4697 was one of the few X-ray faint early-type galaxies for which a long
{\it ROSAT} HRI observation (RH600825A01) exists in the HEASARC archive
(78,744 seconds). The observation did not contain any time intervals with
excessively high background so the entire observation was used. The inner
$8^{\prime} \times 8^{\prime}$ of the HRI image is shown in
Figure~\ref{fig:hri}. A contour map of the HRI image is shown overlaying the
the Digital Sky Survey optical image in Figure~\ref{fig:hri_opt}.
Several X-ray point sources are evident from the contour plot.
Background-subtracted count rates were calculated for each point source
detected within the same $4^{\prime}$ circle used to derive the PSPC and
{\it ASCA} spectra. Background was selected from an annular ring with inner and
outer radius of $4^{\prime}$ and $6^{\prime}$, and was corrected for vignetting
before being subtracted from the source. 
In all, 12 point sources were detected at a significance of 2 $\sigma$ or
higher, corresponding to a detection flux limit of $1.0 \times 10^{-14}$ ergs
s$^{-1}$ cm$^{-2}$ and a limiting luminosity of $3.0 \times 10^{38}$ ergs
s$^{-1}$. From the source counts catalog of a deep HRI 
observation by Hasinger et al.\ (1998), we would expect there to be
$\le$1 serendipitous source in a $4^{\prime}$ circle field at this flux
level or higher. Thus, it is likely that all or nearly all of the X-ray
sources in the field belong to NGC~4697.

The positions, number of counts, significance of detection, and luminosities
of the 12 point sources are shown in Table~\ref{tab:sources}. The count
rates were converted to luminosities using the two-component spectral model
found in \S~\ref{sec:spectral}. Two of the sources (Sources 6 and 11) are
coincident within the position errors with faint optical point sources with
magnitudes of approximately 18, that do not correspond
to any QSO found in the Veron-Cetty \& Veron (1998) catalog. The optical source
near Source 11 has been identified as a bright globular cluster of NGC~4697
(Hanes 1977).
The central source in Figures 1 and 2 appears elongated in the north-south
direction.
When the center is viewed at higher resolution, the source
is split into two sources (Sources 7 and 8 in Table 1), which are
separated by $~\sim7^{\prime\prime}$,
close to the resolution limit of the HRI.
Source 7 is within $2^{\prime\prime}$ of the optical center of
NGC~4697 (R.A. = 12:48:35.71 and Dec.\ = -5:48:02.9), as given
by Wegner et al.\ (1996).
Given the the HRI positional uncertainty of $\sim$5$^{\prime\prime}$,
the additional uncertainty due to the crowding of Sources 7 and
8 (which might contain additional components), and the optical
position uncertainty of about $1\farcs25$, it is possible that
Source 7 or even 8 might be coincident with the optical center of the
galaxy.
Thus, one of these sources might be due to an active nucleus in the
galaxy.
On the other hand, there is no evidence for an AGN in NGC~4697;
for example, the nucleus is not a radio source
(e.g., the NVSS survey, Condon et al.\ 1998). The positive detection
of only 12 sources does not allow us to determine if the sources follow
the de Vaucouleurs stellar distribution, especially considering that one if not
more of the sources is associated with a globular cluster of NGC~4697,
and crowding in the center will underestimate the number of detected
point sources in this region.

What is not evident from Figure~\ref{fig:hri_opt} is the presence of very
faint unresolved X-ray emission within $4^{\prime}$ (the contours of
Figure~\ref{fig:hri_opt} were chosen to highlight the position of the X-ray
point sources). Figure~\ref{fig:hri_diffuse} shows the contours from a more
heavily smoothed HRI image once again overlaying the optical image. The
emission appears elongated in the direction of the optical major axis
of the galaxy. The surface brightness profile of the X-ray emission follows
the optical surface brightness profile derived by
Jedrzejewski, Davies, \& Illingworth (1987) out to 4$^{\prime}$, suggesting
that the X-ray emission is distributed like the stars. However, it was found
that a $\beta$-model profile with a shallow slope ($\beta = 0.4-0.45$) also
fit the data adequately. The faint unresolved emission is actually the
dominant source of emission in the HRI image, comprising
79\% of the total emission. The nature of this
unresolved emission is discussed next.

\section{Discussion} \label{sec:discussion}

The 12 point sources detected within $4^{\prime}$ of NGC~4697 constitute
21
to whether the remaining unresolved emission is the summed emission from
LMXBs below the detection threshold of the HRI or is a low temperature
ISM with $kT\sim0.25$ keV. The absorbed flux of the hard component found
from the spectral fitting of the PSPC and {\it ASCA} data sets
a lower limit on the contribution of LMXBs to the unresolved emission.
In the {\it ROSAT} band, the hard component contributes 40\% of the total
(unabsorbed) flux. Thus, even if LMXBs possess only a hard X-ray component,
another
$\sim20\%$ of the total emission must result from LMXBs below the detection
threshold of the HRI.

As mentioned in the Introduction, LMXBs might also possess a soft X-ray
component. This component would have gone unnoticed in
most Galactic disk or bulge LMXBs,
where the high Galactic column densities associated with the Galactic plane
would completely absorb any very soft emission. Although there does not
appear to be a soft component towards Galactic globular cluster LMXBs that
lie in directions of low column densities, this might be the effect of the
low metallicities in which the LMXBs formed (see Irwin \& Bregman 1999b).
The bulge of M31, however, provides the nearest opportunity to study a sample
of LMXBs in an environment most similar to that found in early-type galaxies.
Supper et al.\ (1997) analyzed the {\it ROSAT} PSPC image of M31 and
found 22 point sources within the inner $5^{\prime}$ of the bulge with
luminosities in the range $10^{36}-10^{38}$ erg s$^{-1}$. Individually, these
point sources had X-ray colors that indicated the presence of a significant
soft component. Seven of the 22 sources had enough counts for
Supper et al.\ (1997) to perform spectral fitting. The best-fit bremsstrahlung
temperatures ranged from 0.45--1.5 keV, very similar to the result found
for NGC~4697 for the same model. Although such a simplistic model is not
physically plausible, it did indicate that the spectrum of the LMXBs were
significantly softer than the canonical temperature of 5--10 keV previously
assumed for LMXBs.

We can search for such a soft component in the point sources suspected to
be LMXBs in NGC~4697. Of the 12 point sources detected by the HRI, sources
1, 6, 10, 11, and 12 were also detected and resolved in the PSPC (sources 10 
and 12 were marginally resolved from each other in the PSPC). As mentioned
above, Source 6 might be associated with background/foreground objects.
The counts from the other four sources were summed in three energy bands, and
two X-ray colors (C21 and C32) were defined from the ratio of the three bands:
\begin{equation} \label{eq:c21}
{\rm C21} =
\frac{\rm counts~in~PI~bins~52-90}{{\rm counts~in~PI~bins~11-41}}
\, ,
\end{equation}
and
\begin{equation} \label{eq:c32}
{\rm C32} =
\frac{\rm counts~in~PI~bins~91-202}{{\rm counts~in~PI~bins~52-90}}
\, .
\end{equation}
Note that these colors are not corrected for absorption.

Sources 1+10+11+12 combined yielded 359.5 background-subtracted counts, or 14\%
of the total X-ray emission within $4^{\prime}$ in the PSPC image.
The colors for the sources were (C21,C32 $ = 0.57 \pm 0.09, 0.94 \pm  0.16)$.
As a comparison, the colors for all emission within $4^{\prime}$ were
(C21,C32 $ = 0.68 \pm 0.05, 0.77 \pm  0.05)$. Thus, the colors of the
four resolved sources are consistent with the colors for the integrated
emission from the galaxy. This argues that the four LMXBs must also possess
a soft component similar to that found in the global spectrum. Otherwise, the
integrated emission would have had significantly different colors than the
point sources. Conversely, the colors predicted from a 5.2 keV
bremsstrahlung model with an absorbing column density of $2.12 \times 10^{20}$
cm$^{-2}$ were (C21,C32 $=1.204,1.893)$. This differs from the colors of
the four point sources by 7.0$\sigma$ and 6.0$\sigma$ for C21 and C32,
respectively. Clearly, if LMXBs were described by only a hard component, the
colors of the four point sources would be significantly harder than what
they are.

\begin{figure*}[htb]
\vskip4.5truein
\includegraphics{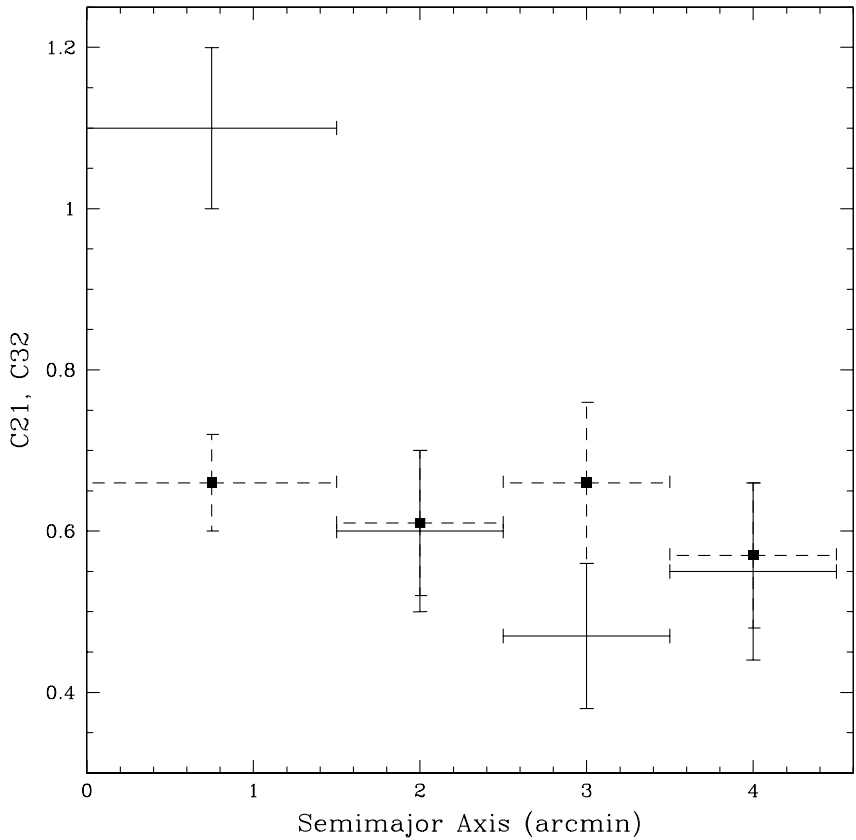}
\caption[]{
C21 (dotted lines) and C32 (solid lines) colors derived from the PSPC data
as a function of semimajor axis for NGC~4697.
\label{fig:profile}}
\end{figure*}

\begin{figure*}[htb]
\vskip4.5truein
\includegraphics{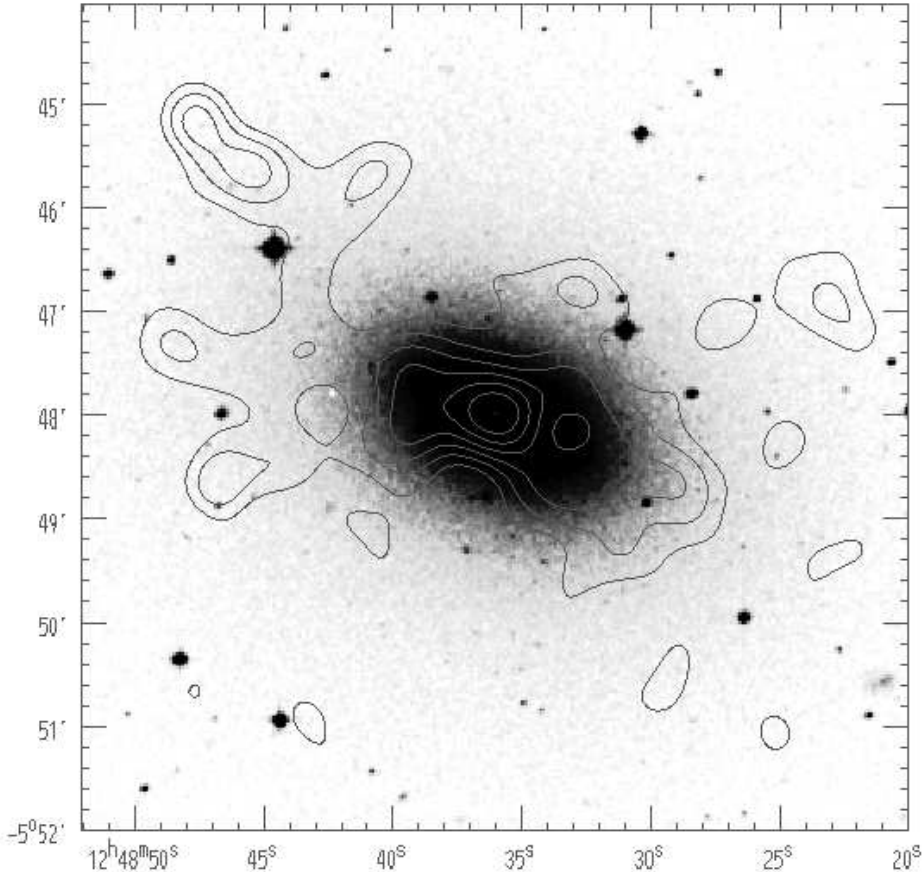}
\caption[]{
{\it ROSAT} HRI contour map overlaid the Digital Sky Survey image of NGC~4697.
The HRI map has been smoothed with a Gaussian of $15^{\prime\prime}$, and the
contours have been chosen to highlight the unresolved emission.
\label{fig:hri_diffuse}}
\end{figure*}

We have calculated the colors of NGC~4697 in elliptical annuli (with
ellipticity of 0.42 and a position angle of 67$^\circ$ to match the optical
profile; Figure~\ref{fig:profile}). C32 peaks inside of $1\farcm5$ but flattens
at larger radii, while C21 is constant throughout the galaxy. The peak in
C32 may result from the presence of an AGN in NGC~4697. The constancy of C21
and C32 outside of $1\farcm5$ implies a single emission mechanism (or two
emission mechanisms with the same spatial distribution), with
colors around 0.6 for both colors, in rough agreement with the colors
of the four point sources. The C32 color of the diffuse emission is somewhat
lower than the C32 value of the four point sources, possibly indicating
the presence of an ISM component, although the discrepancy between the two
colors is only at the 2$\sigma$ level. The colors predicted from a 0.2 keV,
20\% metallicity ISM are (C21,C32 $= 0.54, 0.22)$, whereas the colors for a
0.3 keV, 20\% metallicity ISM are (C21,C32 $= 1.17, 0.46)$. Thus, if any
ISM is present in NGC~4697, its temperature must be below 0.3 keV and at a
low metallicity, or else the C21 color of the LMXBs+ISM would be higher than
the LMXBs alone, which is not observed. In conclusion, we cannot rule out
the presence of some low temperature ISM in NGC~4697, although it is certain
that an ISM cannot constitute a majority of the emission.

Irwin \& Bregman (1999b) found that Galactic and M31 globular cluster LMXB X-ray
colors were correlated with the metallicity of the globular cluster, in the
sense that higher metallicity globular clusters had LMXBs with softer X-ray
colors. If this correlation extended to all LMXBs, it would predict that the
X-ray colors of NGC~4697
should harden with increasing radius, since metallicity decreases with
radius in elliptical galaxies. Furthermore, the metallicity of NGC~4697
is rather low. Within half an effective radius, the average metallicity
is only 50\% solar (Trager et al.\ 2000). The metallicity-color relation of
Irwin \& Bregman (1999b) would predict a C32 color of about 1.5 within
half an effective radius, and an increase with increasing radius. The colors
of the four resolved sources as well as the unresolved emission are at odds
with this metallicity-color relation. Apparently, if such a metallicity-color
relation truly exists for LMXBs, it only applies to LMXBs that reside in
globular clusters. The X-ray source in NGC~4697 associated with a globular
cluster (Source 11) has a C32 color of $1.14 \pm 0.30$, which would be
consistent with the metallicity-color relation if the globular cluster has
a high metallicity.

We also investigated the possibility that LMXBs below the detection
threshold of the HRI could account for the unresolved emission given a
reasonable LMXB luminosity distribution function for NGC~4697.
We assumed a luminosity distribution function $N(>L_X) \propto L_X^{-1.3}$,
which is consistent within the errors with the luminosity distribution
function of point sources in M31 with luminosities greater than
$2 \times 10^{37}$ ergs s$^{-1}$ (Primini et al.\ 1993). The function was
normalized to yield the observed X-ray luminosity of NGC~4697 when integrated
over all LMXB luminosities. This model predicted 11 sources with luminosities
over $3 \times 10^{38}$ ergs s$^{-1}$,
which contributed 18\% to the total X-ray emission from LMXBs. This agrees
well with what was observed with the HRI; neglecting the point source
associated with an unidentified optical counterpart, the remaining 11
detected sources
comprised 19\% of the total emission. Thus, if the luminosity distribution
function of LMXBs of NGC~4697 is similar to that of the brighter LMXBs in
M31, the integrated emission from LMXBs below the detection threshold of the
HRI can account for most of the unresolved emission.

Interestingly, the detection limit of the HRI observation of
$3 \times 10^{38}$ ergs s$^{-1}$ lies above the
Eddington luminosity limit for a $1.4~M_{\odot}$ neutron star. LMXBs
of similar luminosities as the ones found here exist in our own Galaxy.
A compilation by Christian \& Swank (1997) found eight galactic LMXBs with
luminosities greater than $3 \times 10^{38}$ ergs s$^{-1}$
(we have converted their 0.7--4.5 keV luminosities to 0.25-10 keV luminosities
using the spectral model of \S~\ref{sec:spectral}). These high luminosities
imply either that the compact object within the binary is a black hole
(with $M_{BH} \ge 6~M_{\odot}$ for the most luminous binaries) or that the
luminosities truly exceed the Eddington limit for a neutron star. The former
would imply that active binaries with massive black holes are fairly
common in galaxies. It should be noted that the bulge of
M31 lacks the very high luminosity LMXBs that NGC~4697 has; the brightest
LMXB in M31 was only $1.8 \times 10^{38}$ ergs s$^{-1}$ (Supper et al.\ 1997).
However, this is likely the result of small number statistics. Simulations of
the bulge of M31 using a luminosity distribution function of the form
$N(>L_X) \propto L_X^{-1.3}$ indicated that only 1--3
sources with luminosities exceeding $10^{38}$ ergs s$^{-1}$ should be found
in M31. In five separate simulations of M31, the peak luminosity for an
LMXB did not exceed $3 \times 10^{38}$ ergs s$^{-1}$, in agreement with
observation.

We cannot rule out the presence of at least some interstellar medium
in NGC~4697 and X-ray faint early-type galaxies in general. Using the
X-ray temperature--optical velocity dispersion relation of Davis \& White
(1996), any ISM present in NGC~4697 would be expected to have a temperature
around 0.3 keV. This would be very difficult to distinguish from the soft
component from LMXBs on a spectroscopic basis alone. What is needed to
separate the ISM component from the LMXB emission is the high spatial
resolution that can be afforded by {\it Chandra}. Below, we present a
simulation of what we expect the emission from NGC~4697 to look like in the
event that the emission is composed solely of LMXBs.

\begin{figure*}[htb]
\vskip4.5truein
\includegraphics{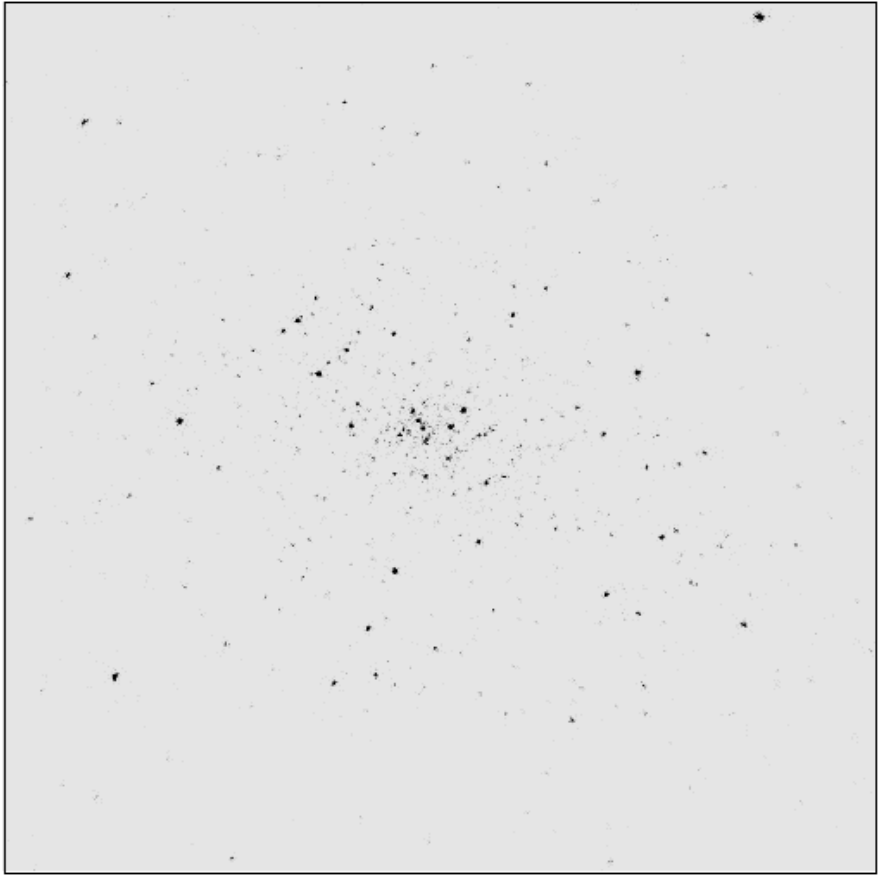}
\caption[]{
A 40,000 s {\it Chandra} simulation of the inner
$4^{\prime} \times 4^{\prime}$ of NGC~4697 in the 0.25--10 keV range, assuming
all of the X-ray emission from NGC~4697 is from LMXBs.
\label{fig:chandra}}
\end{figure*}

\section{Simulation of {\it Chandra} Observation of NGC~4697}
\label{sec:chandra}

We have an approved Cycle 1 40,000 s {\it Chandra} observation of
NGC~4697;
here we show that this observation should resolve the hard and
soft X-ray emission into individual sources, assuming that the
emission is from LMXBs.
Conversely, the observation should cleanly separate a truly
diffuse emission from that of LMXBs.
Since the main goal is to resolve the issue of the very soft component, the
soft X-ray sensitive backside-illuminated (BI) S3 chip of the ACIS-S array
will be used for the observation.
We have used the MARX (Model of {\it AXAF} Response to X-rays;
Wise, Huenemoerder, \& Davis 1997) Simulator to generate a synthetic image
of NGC~4697. The MARX Simulator takes as input the desired
spectral model and spatial distribution model of an X-ray source and creates
an image of the source as it would appear once having passed through
the optics of {\it Chandra}.
The spectral and spatial distribution models described
below were fed into MARX using the ACIS-S BI response to produce an image
of NGC~4697 for a 40,000 s observation.

For the spectra of the LMXBs, we assume a model that best fit the joint
{\it ROSAT} PSPC + {\it ASCA} spectrum of NGC~4697 discussed in
\S~\ref{sec:spectral}.
For the spatial distribution of the X-ray emission, we
assume that the LMXBs have the same spatial distribution as the stellar light.
We take the optical distribution to be a de Vaucouleurs profile
(de Vaucouleurs et al.\ 1991)
with a mean half-light radius of $72^{\prime\prime}$,
an effective semimajor axis of $95^{\prime\prime}$,
an effective semiminor axis of $55^{\prime\prime}$,
resulting in an ellipticity of 0.42, and elongated at
a position angle of 67$^\circ$.
(Jedrzejewski et al.\ 1987;
Faber et al.\ 1989;
Peletier et al.\ 1990).
As before, we assume a luminosity distribution
function for the LMXBs of the form $N(>L_X) \propto L_X^{-1.3}$. The
luminosity distribution function contained point sources with 0.25--10.0 keV
luminosities between $10^{37}$ and $10^{39}$ ergs s$^{-1}$.
A model galaxy was created by drawing LMXBs at random from the luminosity
distribution function and the spatial distribution until the sum of the
luminosities of the LMXBs totaled the X-ray luminosity of the galaxy.
We did not use the observed positions and fluxes of the sources in
Table 1 in the model; the sources are randomly drawn from the
optical surface brightness distribution and X-ray luminosity distribution.
We did not include background in this simulation.
The background in the {\it Chandra} ACIS S-3 chip is highly variable
(Markevitch 1999),
and its level during the observation will affect the ability to detect the
weakest sources.

Figure~\ref{fig:chandra} shows the simulated image of NGC~4697. The image
shows the inner $4^{\prime}$ by $4^{\prime}$ of the galaxy. The excellent
spatial resolution of {\it Chandra} is evident.
The dynamic range of the source brightness in this image is not clear from
this greyscale representation, since all of the strong sources are nearly
the same size (set by the resolution) and black.  However, the sources
cover a range of $\sim$25 in flux.
For our simulation, we assumed that 16 source counts would be needed
to give a $3\sigma$ detection considering the variable background of the
ACIS S-3 chip. This assumption predicts that $\sim$100 sources
would be detected at $\ge3\sigma$ with a detection threshold of
$6 \times 10^{37}$ ergs s$^{-1}$. These 100 sources comprise 50\% of the total
luminosity of the galaxy. In the 0.25--0.8 keV band, $\sim$50 sources were
detected at $\ge3\sigma$.

We have also simulated a {\it Chandra} image where the LMXBs have only a hard
(5.2 keV) bremsstrahlung component that comprises 59\% of the 0.25--10 keV
emission (this was the relative contribution of the hard component
to the total emission found in \S~\ref{sec:spectral}). In this case,
only about 12 LMXBs were detected in the 0.25--0.8 keV band. If LMXBs do possess
a soft component, it will be immediately obvious from the 0.25--0.8 keV
{\it Chandra} image of NGC~4697.

X-ray spectra of some of the brighter sources, X-ray colors of fainter
sources, the cumulative X-ray spectra of individual sources and of
any remaining unresolved diffuse emission can be determined, and
should resolve the question of the origin of the hard and soft
X-ray components in X-ray faint ellipticals.

A number of fundamental questions regarding the X-ray emission from
early-type and Sa galaxies will be answered with {\it Chandra}.
With the expected number of detectable LMXBs, accurate luminosity
distribution functions can be determined. This is something that has been
accomplished for only a handful of nearby galaxies, none of which are
normal early-type galaxies. This will provide important clues to the
stellar evolution of binary stars in galaxies. There is evidence that there
may not be a universal stellar X-ray--to--optical luminosity ratio. A range
of $L_{X, stellar}/L_B$ values has been suggested by comparison of the
X-ray emission of the bulge of M31 to that of Cen A (Turner et al.\ 1997),
and also to several very X-ray faint early-type galaxies
in the Irwin \& Sarazin (1998b) survey. If nearly all the X-ray emission
in NGC~4697 turns out to be stellar in nature, it would be difficult to
explain why the early-type galaxy NGC~5102, for example, has a 0.5-2.0 keV
$L_X/L_B$ value that is 19 times lower than that of NGC~4697
(Irwin \& Sarazin 1998b), if
$L_{X, stellar}/L_B$ is constant from galaxy to galaxy. Without knowledge of
the luminosity distribution function, it cannot be determined if this
difference is a result of a different slope in the distribution function
among galaxies, or if the slopes are the same but the normalizations
of the function (relative to the optical luminosity of the galaxy) are
different. The separation of the stellar and gaseous components is necessary
to resolve this issue.

Since LMXB emission is expected to be present in all early-type galaxies,
the magnitude of this component needs to be known accurately in order to
subtract the LMXB contribution from the total X-ray emission in gas-rich
early-type galaxies. This will make estimates of the mass of elliptical
galaxies based on the assumption that the gas is in hydrostatic equilibrium
more accurate. This is particularly important in the event that LMXBs
possess a strong soft component. Although, the contribution of LMXBs to the
total X-ray emission from very X-ray bright Virgo elliptical galaxies
such as NGC~4472 and NGC~4636 will be small (less than 10\% in the 0.1--2.4
keV band), LMXBs might contribute a significant percentage of the X-ray
emission in galaxies of intermediate X-ray brightness, and will have to be
dealt with accordingly if the true amount of X-ray gas is to be determined.

\section{Conclusions} \label{sec:conclusions}

We have analyzed deep {\it ASCA}, {\it ROSAT} PSPC, and {\it ROSAT} HRI
images of the X-ray faint early-type galaxy NGC~4697. Much like other
X-ray faint early-type systems, the spectrum of NGC~4697 is characterized
by hard (5 keV) plus very soft (0.3 keV) emission. Whereas the nature of the
hard emission is generally regarded as the integrated emission from LMXBs,
we have provided additional evidence that much of the soft emission also
emanates from LMXBs. Four of the 12 HRI point sources were resolved by
the PSPC and were found to have soft X-ray colors that were very similar to
those of the galaxy as a whole. These colors were significantly softer
than the colors predicted if only the hard component was attributed to LMXBs.

The 12 point sources detected by the HRI comprised 21\% of the total
X-ray emission within $4^{\prime}$ of NGC~4697. Given a luminosity
distribution function consistent with that for the brighter point sources in
M31, the remaining unresolved emission could emanate solely from LMXBs
below the detection threshold of the observation. However, the presence of
a low temperature ISM could not be completely ruled out. Higher spatial
resolution data afforded by {\it Chandra} should successfully resolve much
of the LMXB emission, as we have shown in our simulations. The determination
of the origin of the soft component will have important implications 
concerning our understanding of the fate of gas lost from stars in galaxies
as well as the X-ray emission mechanism of LMXBs.

\acknowledgments

We thanks the referee, Fabrizio Brighenti, for many useful comments and
suggestions concerning the manuscript.
This research has made use of data obtained through the High Energy
Astrophysics Science Archive Research Center Online Service, provided
by the NASA/Goddard Space Flight Center. The optical image
of NGC~4697 is from the Digital Sky Survey, which were produced at the
Space Telescope Science Institute. The images of these surveys are based on
photographic data obtained using the Oschin Schmidt Telescope on Palomar
Mountain and the UK Schmidt Telescope. J. A. I. was supported by {\it Chandra}
Fellowship grant PF9-10009, awarded through the {\it Chandra} Science Center.
The {\it Chandra} Science Center is operated by the Smithsonian Astrophysical
Observatory for NASA under contract NAS8-39073. C. L. S. was supported in part
by NASA Chandra grant GO0-1019X. J. N. B. was supported by NASA grant
NAG5-3247.

\end{document}